\newcommand{\msun}{M_{\odot}}

\newcommand{\ltsim}{\protect\raisebox{-0.5ex}{$\:\stackrel{\textstyle <}
        {\sim}\:$}}
\newcommand{\gtsim}{\protect\raisebox{-0.5ex}{$\:\stackrel{\textstyle >}
        {\sim}\:$}}

\documentclass{iaus}
\usepackage{graphicx}

\title[Radiation Pressure in Massive Star Formation] 
{Radiation Pressure in Massive Star Formation}

\author[Krumholz, Klein, \& McKee]   
{Mark R. Krumholz$^1$, Richard I. Klein$^{2,3}$ \break
\and Christopher F. McKee$^{1,2}$}

\affiliation{$^1$ Physics Department, University of California,
Berkeley, Berkeley, CA 94720-7304 USA
\break emails: krumholz@astron.berkeley.edu, cmckee@astron.berkeley.edu
\\[\affilskip]
$^2$ Astronomy Department, University of California,
Berkeley, Berkeley, CA 94720-7304 USA
\break
email: klein@astron.berkeley.edu
\\[\affilskip]
$^3$ Lawrence Livermore National Laboratory, Livermore, CA 94550 USA
}

\pubyear{2005}
\volume{227}  
\pagerange{}
\date{}
\jname{Massive Star Birth: A Crossroads of Astrophysics}
\editors{R. Cesaroni, E. Churchwell, M. Felli, \& C.M. Walmsley, eds.}

\begin{document}

\maketitle

\begin{abstract}
Stars with masses of $\gtsim 20$ $\msun$ have short Kelvin times that
enable them to reach the main sequence while still accreting from
their natal clouds. The resulting nuclear burning produces a huge
luminosity and a correspondingly large radiation pressure force on
dust grains in the accreting gas. This effect may limit the upper mass 
of stars that can form by accretion. Indeed, simulations and analytic
calculations to date have been unable to resolve the mystery of how
stars of $50$ $\msun$ and up form. We present two new ideas to solve
the radiation pressure problem. First, we use three-dimensional
radiation hydrodynamic adaptive mesh refinement simulations to study
the collapse of massive cores. We find that in three dimensions a
configuration in which radiation holds up an infalling envelope is
Rayleigh-Taylor unstable, leading radiation driven bubbles to collapse 
and accretion to continue. We also present Monte Carlo radiative
transfer calculations showing that the cavities created by
protostellar winds provides a valve that allow radiation to escape the 
accreting envelope, further reducing the ability of radiation pressure 
to inhibit accretion.
\end{abstract}

\firstsection 
\section{Introduction}

Stars with masses $\gtsim 20$ $\msun$ have short Kelvin times that
enable them to reach the main sequence while still accreting
\cite{shu87}. The resulting nuclear burning produces a huge
luminosity and a correspondingly large radiation pressure
force on dust grains in the incoming gas. This force can
exceed the star's gravitational pull, possibly halting accretion and
setting an upper limit on the star's final mass. Early spherically
symmetric calculations found limits $\sim 20-40$ $\msun$ (\cite[Kahn
1974;]{kahn74} \cite[Wolfire \& Cassinelli 1987]{wolfire87})
for typical galactic metallicities. More recent
non-spherical calculations have loosened that constraint by
considering the role of accretion disks (\cite[Nakano 1989;]{nakano89}
 \cite[Nakano, Hasegawa, \& Norman 1995;]{nakano95} \cite[Jijina \& Adams
1996]{jijina96}), which concentrate the incoming matter into a
smaller solid angle and also cast a shadow of reduced 
radiation pressure. Even with the effects of a disk, however,
radiation pressure can still be a significant barrier to
accretion. Numerical simulations by \cite[Yorke \& Sonnhalter
(2002)]{yorke02} found limiting masses of $\sim 40$ $\msun$ before
radiation pressure reversed the inflow. However, observations show
that considerably more massive stars exist, and their formation
mechanism remains uncertain.

In this paper we present two new mechanisms that allow accretion onto
massive stars to overcome the radiation pressure barrier and continue
to higher masses. First, we perform 3D self-gravitational radiation
hydrodynamic
adaptive mesh refinement simulations of the collapse of massive
protostellar cores, starting from both quiescent and turbulent initial 
conditions. We find that sufficiently massive stars create
radiation-driven bubbles in accreting flows, but these bubbles
become unstable and collapse due to a radiation Rayleigh-Taylor
instability. We discuss the numerical 
methods we use in our simulations in \S~\ref{code}, and we present
the simulation results in \S~\ref{simresults}. Our second new
mechanism is protostellar winds, which we discuss in
\S~\ref{winds}. Outflows launched from massive stars are
essentially dust-free, and that as a result they are optically thin
below their caps of swept-up material. We present Monte Carlo
calculations of radiative transfer through envelopes with outflow
cavities, and show that the cavities greatly reduce the radiation
pressure on the accreting gas. These two new mechanisms indicate that
radiation pressure is a much less significant barrier to massive star
formation than previous work has suggested.

\section{Simulations}
\label{code}

\subsection{Numerical Methods}

In our simulations, we solve the equations of self-gravitational
radiation hydrodynamics in the gray, flux-limited diffusion
approximation. These are
\begin{eqnarray}
\frac{\partial \rho}{\partial t} + \nabla \cdot (\rho \mathbf{v}) & =
& 0 \\
\frac{\partial}{\partial t} (\rho \mathbf{v}) + \nabla \cdot (\rho
\mathbf{v v}) & = & -\nabla P - \rho \nabla \phi + \frac{\kappa_{\rm
R,0}}{c} \mathbf{F} \\
\frac{\partial}{\partial t} (\rho e) + \nabla \cdot \left[(\rho
e+P)\mathbf{v}\right] & = & -\rho \mathbf{v} \cdot \nabla \phi -
\kappa_{\rm P,0} (4 \upi B - c E) - \frac{\kappa_{\rm R,0}}{c}
\mathbf{v} \cdot \mathbf{F} \\
\nabla^2 \phi & = & 4 \upi G \rho \\
\frac{\partial E}{\partial t} + \nabla \cdot \mathbf{F} & = &
\kappa_{\rm P,0} (4 \upi B - c E) + \frac{\kappa_{\rm R,0}}{c} \mathbf{v}
\cdot \mathbf{F} \\
\mathbf{F_0} & = & -\frac{c \lambda(E_0)}{\kappa_{\rm R,0}} \nabla E_0,
\end{eqnarray}
where $\rho$, $\mathbf{v}$, $P$, and $e$ are the density, velocity,
pressure, and specific kinetic plus internal energy of the gas, $\phi$ is the
gravitational potential, $\kappa_{\rm R}$ and $\kappa_{\rm P}$ are the 
Rosseland and Planck mean opacities, $E$ is the radiation energy
density, $\mathbf{F}$ is the radiation flux, and $\lambda(E)$ is the
flux limiter \cite{levermore81}. Subscript zeros indicate that
quantities are evaluated in the frame co-moving with the gas.

The Berkeley adaptive mesh refinement code that we have developed
(\cite[Truelove \etal\ 1998]{truelove98}; \cite[Klein 1999]{klein99};
\cite[Howell \& Greenough 2003]{howell03}) consists of
three modules. The hydrodynamics module updates the state with the
hydrodynamic terms of the equations using a conservative Godunov
scheme that is second-order accurate in both space and time. The gravity
module then solves the Poisson equation for $\phi$ using a multigrid
method. Finally, the radiation module performs a radiation update in
two steps. It handles the radiation diffusion and emission/absorbtion terms
($\nabla\cdot\mathbf{F}$ and $\kappa_{\rm P}(4\pi B-cE)$) with an
implicit method and the
work/advection and force terms ($\kappa_{\rm R} \mathbf{F}/c$ and
$\kappa_{\rm R} \mathbf{v}\cdot \mathbf{F}/c$) with an explicit
method. This approach has the advantage of minimizing the cost of the
implicit solve, yet it does not cause instability because the
diffusion and emission/abosbtion terms are larger than the
work/advection terms by $O(c/v)$. The solvers all operate on adaptive
grids (\cite[Berger  \& Oliger 1984]{berger84}; \cite[Berger \& Collela
1989]{berger89}; \cite[Bell \etal\ 1994]{bell94}), allowing the
calculations to reach an effective linear resolution up to 16,384.

We represent any stars that form in the simulation using the
Lagrangian sink particle algorithm of \cite[Krumholz, McKee, \& Klein
(2004)]{krumholz04}. The sink particles can move through the grid,
accrete mass, and interact gravitationally with the gas. They also act
as sources of radiation. We couple each sink particle to a
protostellar model based on \cite[McKee \& Tan (2003)]{mckee03}, that
updates the mass, radius, luminosity, and internal state of a star
based on the accretion rate determined by the simulation. The model
includes luminosity from accretion, Kelvin-Helmholtz contraction,
deuterium burning, and hydrogen burning.

\subsection{Initial Conditions}

We simulate the formation of stars from both turbulent and quiescent
massive pre-stellar cores. For quiescent runs we begin our simulations
with spheres of gas of $100$ $\msun$ or $200$ $\msun$ with a density
profile $\rho\propto r^{-2}$,
a radius of $0.1$ pc or $0.2$ pc, and temperature 40 K, in solid body
rotation with a ratio of rotational kinetic energy to
gravitational potential energy $\beta=0.02$, the median observed for low
mass cores \cite{goodman93}. The computational domain is a box 0.4 pc
or 0.6 pc on a side for these runs, and the maximum resolution of
these runs is $10$ AU and $14$ AU, respectively. Note that these
initial conditions are very similar to those used by \cite[Yorke \&
Sonnhalter (2002)]{yorke02}.

For the turbulent initial conditions, we simulate a
core based on the \cite[McKee \& Tan (2003)]{mckee03} model. We use a
density profile $\rho\propto r^{-2}$ continuing self-similarly to
the edge of the computational domain, a box $0.8$ pc accross. The
density profiel is normalized so that $150$ $\msun$ of gas lies within
$0.1$ pc of the origin. We impose a Gaussian-random turbulent velocity
field with its power on wavenumbers of $k=1-2$, where $k=1$
corresponds to a wavelength equal to the size of the computational
domain. The Mach number $\mathcal{M}$ of the turbulence is chosen so that the
kinetic and potential energies within the computational domain are
equal, giving $\mathcal{M}=8.6$. The maximum resolution is $10$ AU.

In our runs we refine any cell
that is within 32 cells of a sink particle, where the gradient of
radiation energy density satisfies $\left|\nabla E\right|/E>0.25$, or
where the cell violates the Jeans condition for self-gravitational
stability \cite{truelove97}.

\section{Simulation Results}
\label{simresults}

\subsection{Quiescent Initial Conditions}

The results of the $100$ $\msun$ and $200$ $\msun$ runs are
qualitatively very similar, so for simplicity we only discuss them
together. The accretion process appears to pass through three
phases. During the first phase, when $M\ltsim 17$ $\msun$, the
accretion flow is smooth and there are no obvious effects from
radiation pressure. Low angular momentum gas accretes directly onto
the star, and higher angular momentum gas falls into a Keplerian
accretion disk. At $M\ltsim 5$ $\msun$ the accretion luminosity is
dominant, but thereafter it becomes a relatively minor
component of the total energy output.

\begin{figure}
\centerline{
\scalebox{0.8}{\includegraphics{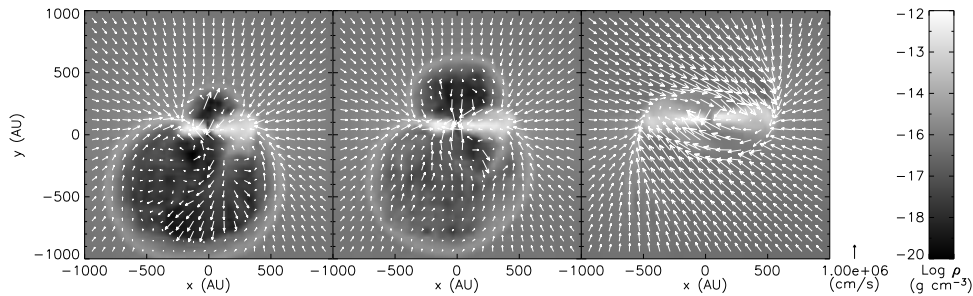}}
}
\caption{
\label{bubblephase}
The plot shows a slice in the XZ plane at three
different times, showing the initial growth, instability, and collapse
of a radiation bubble. The times of the three slices are $1.5\times 10^4$,
$1.65\times 10^4$, and $2.0\times 10^4$ yrs, and the stellar masses are
$21.3\msun$, $22.4\msun$, and $25.7\msun$.}
\end{figure}

The second phase begins at $\sim 17$ $\msun$, when the star is
luminous enough for radiation pressure to start repelling gas. This
leads to the formation of radiation-driven bubbles above and below the 
accretion disk, as shown in Figure \ref{bubblephase}, first panel. The
bubbles are asymmetric above and below the accretion
disk and with respect to the polar axis, an effect that earlier
single-quadrant 2D simulations could not model. The asymmetry is the
result of a positive feedback loop. If the radiation bubble has
expanded farthest in a certain direction, then there is less gas
above the star in that direction. As a result, that
direction represents the path of minimum optical depth out of the
core and radiation is collimated towards it. Since more radiation is
flowing in that direction, there
is a greater radiation pressure force, which in turn causes the bubble 
to expand even faster. At no point
during the bubble phase does accretion stop, although it does become
clumpy and erratic. Infalling gas that reaches a radiation bubble
travels around its edge until it reaches the equatorial plane and the
accretion disk. The disk itself is unaffected by the radiation
pressure due to its high optical depth. This phase lasts $\sim 10^4$
yr.

The transition to the third phase of accretion happens at a mass of
$22-26$ $\msun$, when the bubble becomes unstable and begins to
collapse. The collapse of the bubble appears to occur when the bubble
walls stop accelerating away from the star, so that there is a net
gravitational force towards the star within the walls. The instability 
leads to ripples in the bubble walls, and eventually to gas breaking
through into the bubble. This is shown in Figure \ref{bubblephase},
second panel. Physically, this process is easy to understand as
Rayleigh-Taylor instability: the gas is a heavy fluid, and radiation
is in effect a light fluid. Once the local gravity is such that the
radiation is below the gas, instability naturally results. However,
the instability seems to begin with a large wavelength mode passing
through the pole, in which the entire shell ripples. This mode
cannot exist in an axisymmetric simulation, which is probably why
previous work has not seen it.

Once enough gas has flooded in, the bubble effectively
disappears, What is left is a remnant of the bubble wall. After this
point, gas that falls towards the star is deflected to the side by the 
radiation field until it reaches the dense wall. The gas then flows
along the wall into the accretion disk, and is able to accrete. This
configuration is shown in Figure \ref{bubblephase}, third panel. To
the point we have evolved the simulations thus far, this appears to be 
the final configuration of the system. However, it is possible that
there may be more episodes of bubble formation and destruction. At the 
time of this writing, the simulations have reached a mass of $> 27$ 
$\msun$, considerably larger than the $22.9$ $\msun$ \cite[Yorke \&
Sonnhalter (2002)]{yorke02} found for their calculations with gray
radiative transfer. Our simulations are continuing to determine an
upper mass limit.

\subsection{Turbulent Initial Conditions}

Our turbulent run shows results
qualitatively quite similar to our queiscent runs. In the early stages 
of the simulation, a star with a disk around it forms. The orientation 
of the disk varies in time, as the angular momentum of the accreting
gas changes in time due to the turbulence. At $17-18$ $\msun$, a
bubble grows around the star. We 
have evolved the star to a mass of $\sim 25$ $\msun$ as of this
writing. The star is still in the bubble phase, and the bubble shows
no signs of collapsing yet, as it showed by this mass in the quiescent
cases. The bubbles above and below the accretion disk have also grown
to a diameters of almost $2000$ AU, larger than they became in the
quiescent case. We believe the changes in bubble evolution are
probably due to a combination of the change in the normalization of
the density profile ($150$ $\msun$ versus $100$ $\msun$ within 0.1 pc)
and to the turbulence, both of which reduce the effective radial
infall velocity at the start of the bubble phase and thus allow the
expansion of the bubble to accelerate for longer times. The mean
accretion rate in the simulation thus far is $2.8\times 10^{-3}$
$\msun$ yr$^{-1}$. Because our initial conditions are not in
virial balance on small scales, this is slightly larger than the $1.6\times
10^{-3}$ $\msun$ yr$^{-1}$ that \cite[McKee \& Tan
(2003)]{mckee003} predict for our initial density profile.

We do not see any evidence of fragmentation to form many stars, as
found by \cite[Dobbs, Bonnell, \& Clark (2005)]{dobbs05} in
simulations with fairly similar initial conditions. We hypothesize
that the different outcome stems from our more detailed treatment of the
radiation. Dobbs \etal\ use either an isothermal or a
barotropic equation of state, and find that the barotropic equation of
state reduces their number of fragments from tens to a few. This
points to the importance of heating in determining whether fragmentation
occurs. However, even a barotropic equation of state misses the
vast majority of the heating. Since neither our simulations nor those
of Dobbs \etal\ resolve the protostellar radius, most of the energy
released by the infalling gas is released at sub-grid scales, in the
final plunge onto the stellar surface. Our method
captures this energy, by including the accretion luminosity in the energy 
output by our sink particle, but a barotropic equation of state omits
it. In our simulations, the
high accretion luminosity that results from the high pressure in the
protostellar core heats the gas within $\sim 1000$ AU of the star
(where Dobbs \etal\ get fragmentation) to hundreds of K in
only a thousand years. This is shorter than the free-fall time
of $\sim 4000$ yr at 1000 AU, so the heating prevents fragmentation.

\section{Protostellar Outflows}
\label{winds}

We also consider the effects of protostellar outflows on the
accretion process (\cite[Krumholz, McKee, \& Klein 2005]{krumholz05}
for a detailed discussion). \cite[Beuther, Schilke, \&
Geuth (2004)]{beuther04} report interferometric measurements
showing outflows from massive protostars with collimation
factors of $\sim 2$ up to $\sim 10$. They conclude that high and low
mass outflows have similar collimation factors, typically
$2-5$. (The collimation factor is the ratio of the outflow's length to
its width.) \cite[Richer \etal\ (2000)]{richer00} show that the
momentum of CO outflows driven by massive stars scales with the
bolometric luminosity of the source in the same manner as for low mass
stars. From these observations, the natural conclusion is that low and
high mass stellar outflows have a common driving mechanism and similar
morphologies.

This analysis suggests that massive protostellar outflows originate
near their source stars, inside the stars' dust destruction radius. As 
a result, they should be dust-free when launched. Once the outflowing
gas is far enough from the star, grains can begin to grow. However,
the wind density is relatively low and its speed is very high. For a
range of possible shapes of the cavity filled by outflowing gas,
\cite[Krumholz \etal]{krumholz05}
show that the largest grains ejected in the wind from a $50$ $\msun$
star at a typical velocity of 1000 km s$^{-1}$ reach sizes of no more
than $\sim 10^{-4}$ $\mu$m before escaping the accreting core. Since
the optical depth of grains to radiation both from the star and the
envelope obeys $\kappa\propto r_{\rm gr}$, this means that the outflow 
channel will be quite optically thin up to its cap of swept-up
material. Since the surrounding envelope is very optically thick, we
expect the outflow cavity to collimate the radiation and carry it away 
from the accreting gas.

To see how this affects radiation pressure, we compute radiative
transfer through envelopes witih and without outflow cavities using
the Monte Carlo code of \cite[Whitney \etal(2003a,b)]{whitney03a}, and 
then compute radiation pressure forces within the envelope using a ray 
tracing code. We find that even in a very low surface density envelope 
with $\Sigma\approx 0.1$ g cm$^{-3}$, an outflow cavity reduces the
radiation pressure force in the equatorial plane by a factor of $\sim
5$. At the higher surface densities typical of most massive star
forming regions, the collimation is even more effective, and the
radiation pressure is reduced by a factor of $\gtsim 10$. Since
observations show that such outflows are ubiquitous, we expect suggest 
that radiation pressure is a much less significant barrier to massive
star formation than calculations without protostellar outflows have
found.

\section{Conclusions}

Massive star formation by accretion is much easier than previous work
suggested. In three dimensions, radiation holding up
accreting gas is subject to Rayleigh-Taylor instability. Radiation
deflects gas into optically thick fingers that accrete onto the star,
while the radiation diffuses out around the compressed
gas. Differing initial conditions, including ones with turbulence, do
not appear to change this basic evolution. Furthermore, even without
this instability, the outflows from massive stars are by themselves
very effective at collimating radiation and carrying it away from the
accreting gas. Outflows cavities produce order-of-magnitude reductions 
in the radiation pressure force experienced by inflowing gas, and thus 
make accretion even easier. We therefore conclude that radiation
pressure cannot  prevent the formation of massive stars by accretion.

\begin{acknowledgements}
This work received financial and computing support from: NASA GSRP
grant NGT 2-52278 (MRK); the US Department of Energy at the Lawrence
Livermore National Laboratory under the auspices of contract
W-7405-Eng-48 (RIK and MRK); NASA ATP grant NAG 5-12042 (RIK and CFM);
NSF grant AST-0098365 (CFM); the NSF San Diego Supercomputer Center
through NPACI program grant UCB267 (all); and the National Energy Research
Scientific Computing Center, which is supported by the Office of
Science of the U.S. Department of Energy under Contract
No. DE-AC03-76SF00098, through ERCAP grant 80325 (all).
\end{acknowledgements}

\end{document}